\documentclass[aps,showpacs,twocolumn,groupedaddress]{revtex4-1}
\usepackage[centertags]{amsmath}
\usepackage{amsfonts}
\usepackage{epstopdf}
\usepackage{color}
\usepackage{graphicx}
\usepackage{dcolumn}
\usepackage{bm}

\begin{document}


\title{Three-wave resonant interactions in the diatomic chain with cubic anharmonic potential: theory and
simulations}

\author{A. Pezzi$^{1}$, G. Deng$^{2}$,  Y. Lvov$^{3}$, M. Lorenzo$^{1}$, M. Onorato$^{1,4}$}
\affiliation{$^1$ Dipartimento di Fisica, Universit\`a degli Studi di Torino, 10125 Torino, Italy}
\affiliation{$^2$Department of Mathematics and Statistics, Macquire University, Sydney, New South Wales 2109, Australia}
\affiliation{$^3$ Department of Mathematical Sciences, Rensselaer Polytechnic Institute, Troy, New York 12180, USA}
\affiliation{$^4$ Istituto Nazionale di Fisica Nucleare, INFN, Sezione di Torino, 10125 Torino, Italy}

\begin{abstract}
We consider a diatomic chain characterized by a cubic anharmonic potential. After diagonalizing the harmonic case,
we study in the new canonical variables,  the nonlinear interactions between the acoustical and optical branches of the dispersion relation.  Using the  {\it wave turbulence} approach, we formally derive two coupled wave kinetic equations, each describing
the evolution of the wave action spectral density associated to each branch.
An $H$-theorem shows that there exist an irreversible transfer of energy that leads to an equilibrium solution characterized by the equipartition of energy in the new variables. While in the monoatomic cubic chain, in the large box limit, the main nonlinear transfer mechanism is based on four-wave resonant interactions, the diatomic one is ruled by a three wave resonant process (two  acoustical and  one optical wave): thermalization happens on shorter  time scale for the diatomic chain with respect to the standard chain. Resonances are possible only if the ratio between the heavy and light masses is less than 3.  Numerical simulations of the deterministic equations support our theoretical findings.
\end{abstract}

\maketitle

\section{\label{intro}INTRODUCTION}

Relaxation and thermalization in one dimensional chains are  important research topics in
statistical mechanics and solid state physics \cite{lepri2016thermal}. The first important contribution was given
by Fermi and his collaborators in Los Alamos in the early fifties \cite{fermi1955alamos}.
They analysed numerically a  one dimensional monoatomic  chain, including a cubic or quartic
anharmonic potential, the $\alpha$- and $\beta$-FPUT chains, respectively. At that time, the importance of a thermalization, fundamental for establishing a conduction {\it \`a la} Fourier, was already recognized. Linear systems characterized by an harmonic potential do not contain any intrinsic mechanism that leads to the thermalization  and their conduction properties are  anomalous  \cite{lepri2016thermal}. Despite the presence of nonlinearity, in \cite{fermi1955alamos} the thermalization was not found  and the phenomenon of recurrence, typical of integrable systems, was observed. The only reason for this ``partial unsuccess"   has to be found in the lack of a sufficiently powerful computer. Nowadays, modern numerical computations have highlighted the fact that the same initial conditions provided in
\cite{fermi1955alamos}  can lead to a thermalized spectrum, see for example  \cite{Ponno2011}.
Some years later, it has been given the evidence  that, in the large box limit, the mechanism that leads to the thermalization in chains like the $\alpha$- and $\beta$-FPUT is the four-wave resonant interaction process \cite{PNAS,lvov2018double}, see also \cite{pistone2018universal}.  Numerical confirmation of these predictions
can be also found in  \cite{fu2019universal1,fu2019universal}.

In this paper, we consider an $\alpha$-FPUT model but characterized by alternating masses, i.e. a diatomic chain with cubic potential (quadratic nonlinearities in the equation of motion) and we study the properties of thermalization within the wave turbulence  framework \cite{nazarenko2011wave, zakharov2012kolmogorov}. Numerical simulations of a diatomic $\beta$-FPUT chain and of the diatomic Toda lattice were considered in \cite{fu2019nonintegrability} and it was shown that the thermalization time followed the same scaling as the one for monoatomic chains  $\alpha$- and $\beta$-FPUT \cite{onorato2015route,lvov2018double} and the nonlinear Klein-Gordon equation \cite{pistone2018thermalization}. From a mathematical point of view, we point out a rigorous result in \cite{maiocchi2019freezing} for a diatomic chain where it was proved that, in the limit of small temperature and large ratio between the masses,  the exchange of energy between the modes of the optical branch and those of the acoustic one is practically null for the majority of initial conditions up to some time estimated in \cite{maiocchi2019freezing} (see also \cite{galgani1992problem,bambusi1993exponential}). Here our approach, not rigorous but fully supported by numerical computations, leads us to the conclusion that, if the ratio between the large mass and the small one is less than 3, then an exchange of energy between the acoustical and the optical branches can take place. The mechanism responsible for this transfer is a resonance between two acoustical waves and one optical, i.e. a three-wave resonant interaction process. Note that such processes is forbidden in the monoatomic $\alpha$-FPUT system, \cite{bustamante2019exact},  which is ruled by a four-wave one.

The paper is organized as follows: in Section \ref{sec:linearcase} we describe the model, introduce the canonical variables that diagonalize the harmonic hamiltonian and derive the nonlinear equations in those variables. In Section \ref{sec:stat} we introduce the statistical description, derive the two couple kinetic equations with their equilibrium solutions and then in \ref{sec:num}
we verify our findings with numerical simulations. Conclusions follow.

\section{\label{sec:linearcase}The model }

We consider a chain of $2N$ masses connected by springs at a distance $a$ from each other. We denote $M$ the odd masses and $m$ the even masses; whereas their position at rest in the lattice is $x_n=na$, their displacement with respect to the equilibrium position is $y_n(t)$ and $p_n(t)$ is the linear momentum. We assume periodic boundary conditions so that  $y_{2N}=y_0$. Besides standard Hooke forces between neighbouring masses, we include nonlinear forces, i.e. an anharmonic potential. The Hamiltonian takes the following form:

\begin{equation}
\begin{split}
	&H =\sum_{n=0}^{N-1}\frac{p _{2 n+1}^2}{2 M}+\sum_{n=1}^{N} \frac{p_{2 n}^2}{2m}+\\
	&+\frac{\chi}{2} \sum_{n=0}^{2N-1}  (y_{n+1}-y_{n})^2+  \frac{\alpha}{3} \sum_{n=0}^{2N-1}  (y_{n+1}-y_{n})^3,
	 \label{cubicpot1}
\end{split}
\end{equation}
where $\chi$ and $\alpha$ are the coefficients of the harmonic and anharmonic potential, respectively.
The equations of motions can be directly written for $n=1,2,...,N$ as:
\begin{equation}
\begin{split}
&m\ddot{y}_{2n}=\chi(y_{2n+1}+y_{2n-1}-2y_{2n})+\\
&+\alpha \bigl[(y_{2n+1}-y_{2n})^2-(y_{2n}-y_{2n-1})^2 \bigr]
\end{split}
\label{eqmotionnl1}
\end{equation}
and
 \begin{equation}
\begin{split}
&M\ddot{y}_{2n+1}=\chi(y_{2n+2}+y_{2n}-2y_{2n+1})+\\
&+\alpha \bigl[(y_{2n+2}-y_{2n+1})^2-(y_{2n+1}-y_{2n})^2 \bigr].
\label{eqmotionn2}
\end{split}
\end{equation}

\subsection{The linear case}
It is well known that in the linear case the solutions can be looked in the form
\begin{equation}
\begin{split}
&y_{2n}(t) =
	A_k e ^ {i(2naq_k-\omega_k t)} \\
&	y_{2n+1}(t) = B_k e ^{i((2n+1)aq_k-\omega_kt)},
\end{split}
\label{four_ser}
\end{equation}
where $a$ is the lattice spacing, $\omega_k$ is an angular frequency and $q_k$ are  discrete wave numbers defined as:
\begin{equation}
\label{wavenumber}
q_k=\frac{\pi k}{Na} \qquad k \in \bigg(\!-\frac{N}{2},\frac{N}{2}\bigg].
\end{equation}
Inserting (\ref{four_ser}) in the equations of motion, we get the well known acoustic and optical branches of the dispersion relation:
\small{
\begin{equation}
\label{omega2}
\omega^{\pm}_k= \sqrt{\chi \frac{m+M}{mM} \Bigg[1\pm
\sqrt{1-\frac{4mM}{(m+M)^2}\sin^2(aq_k)} \Bigg]},
\end{equation}}
where  $``+''$ indicates the optical branch while $``-''$ indicates the acoustic one.
These useful relations follow:
\begin{equation}
\label{betarelation}
\beta_k^\pm =\frac{B_k}{A_k}=
\frac{2\chi-m\omega^2_\pm(q_k)}{2\chi\cos(aq_k)}=
\frac{2\chi\cos(aq_k)}{2\chi-M\omega^2_\pm(q_k)}
\end{equation}
and $\beta_k^{+}\beta_k^{-}=-m/M$

\subsection{The nonlinear case: normal variables}
The goal of this section is to transform the equations to a form suitable for developing a
statistical theory. The first step consists in diagonalizing the unperturbed Hamiltonian. We introduce the following notation for the Discrete Fourier Transform:
\begin{equation}
\label{positions}
\begin{split}
&Q_k =\frac{1}{{N}}\sum_{n=1}^{N} y_{2n} e^{-i2naq_k},  \\
&R_k =\frac{1}{{N}} \sum_{n=0}^{N-1} y_{2n+1}e^{-i(2n+1)aq_k}
\end{split}
\end{equation}
and
\begin{equation}
\label{moments}
\begin{split}
&P_k =\frac{1}{{N}}\sum_{n=1}^{N} p_{2n} e^{-i2naq_k} \\
&G_k =\frac{1}{{N}} \sum_{n=0}^{N-1} p_{2n+1}e^{-i(2n+1)aq_k}.
\end{split}
\end{equation}
Writing the Hamiltonian (\ref{cubicpot1}) in terms of Fourier variables, we obtain:
\begin{equation}\label{HFourier}
\begin{split}
&H=\sum_{k} \bigg[ \frac{\vert P_k \vert^2}{2m} +
\frac{\vert G_k \vert^2}{2M} +\chi \bigl[ \vert Q_k \vert^2 + \vert R_k \vert^2+\\
&-\cos(aq_k)(Q_kR_{k}^{*}+Q_{k}^{*}R_k)\bigr]\bigg]+{2i\alpha}\sum_{k_2,k_3,k_4}
[(-1)^lQ_2R_3R_4+\\
&+R_2Q_3Q_4]\sin(aq_2)\delta_{2+3+4,0}\;\;,
\end{split}
\end{equation}
where $Q_i=Q_{k_i}$, $R_i=R_{k_i}$, $\delta_{2+3+4,0}=\delta_{k_2+k_3+k_4,0}$ and  $l$ accounts for the periodicity of the Fourier space, so that the Kronecker $\delta$ is equal to 1  when $k_1+k_2+k_3=l N$, with $l=\{0,\pm 1\}$. Being the Fourier series a canonical transformation, then the equations of motion can be written directly as:
\begin{equation}
\label{2.51}
\dot{Q_k}=\frac{\partial H}{\partial P_k^*},\;\;
\dot{P_k}=-\frac{\partial H}{\partial Q_k^*}, \;\;
\dot{R_k}=\frac{\partial H}{\partial G_k^*}\;\;
\dot{G_k}=-\frac{\partial H}{\partial R_k^*}.
\end{equation}
While for the monoatomic chain the quadratic part of the Hamiltonian is diagonalized in Fourier variables, this does not happens for the diatomic case and an extra canonical transformation has to be performed in order to diagonalize it. Using standard tools (see appendix), the system can be diagonalized using the following canonical transformation:
\begin{equation}
\label{canonical_transform}
\begin{split}
&\widetilde{Q}_k^{s}=\frac{m}{\mu_k^s}Q_k+\beta_k^{s}\frac{M}{\mu_k^{s}}R_k	\\
&\widetilde{P}_k^{s}=P_k+\beta_k^{s}G_k
\end{split}
\end{equation}
where $s=+$ or $s=-$, i.e. the optical or the acoustical branch, and
\begin{equation}
\label{e:mu_k}
\mu_k^s=m+(\beta_k^s)^2 M.
\end{equation}
%
%
%
The harmonic part of the Hamiltonian is now given by
\begin{equation}
H_{\mathrm{0}}=
\sum_{k,s}\bigg[\frac{\lvert \widetilde{P}_k^s \rvert^2}{2 \mu_k^s}+
\frac{1}{2} \mu_k^s \omega_s^2(q_k) \lvert \widetilde{Q}_k^s \rvert^2
\biggr],
\end{equation}
and the full Hamiltonian is reported in the appendix, see eq. (\ref{hamiltontransf}).
To apply the {\it wave turbulence description}
\cite{zakharov2012kolmogorov}, it is convenient to introduce the following normal variables:
\begin{equation}
\label{wave-action1}
\begin{split}
a_k^{s}=\frac{i}{\sqrt{2\mu_k^{s}\omega_k^{s}}}
(\widetilde{P}_k^{s}-i\mu_k^{s}\omega_k^{s}\widetilde{Q}_k^{s}),
\end{split}
\end{equation}
where $a_k^{s}$, with $s=+$ or $s=-$, is related to the optical or acoustical branch
and $\omega_k^{s}$ are now taken as the positive branches.
Within these variables, the equations are written  in the following universal form:
\onecolumngrid
\rule{\textwidth}{0.4pt}
\begin{equation}
\begin{split}
&	i\frac{da_1^{+}}{dt}=\omega_1^{+}a_1^{+} +\sum_{2,3}\big\{ [\bar{V}_{1,2,3}^{(1)} a_2^{+} a_3^{+}+\bar{V}_{1,2,3}^{(2)} a_2^{-} a_3^{-}
+ \bar{V}_{1,2,3}^{(3)} a_2^{+} a_3^{-}] \delta_{1,2+3}+
[\bar{V}_{1,-2,-3}^{(1)} a_2^{+*} a_3^{+*}+ \bar{V}_{1,-2,-3}^{(2)} a_2^{-*} a_3^{-*} +
\\&
+ \bar{V}_{1,-2,-3}^{(3)} a_2^{+*} a_3^{-*}]\delta_{1+2+3,0}+
[2\bar{V}_{1,2,-3}^{(1)} a_2^{+} a_3^{+*}+2\bar{V}_{1,2,-3}^{(2)} a_2^{-} a_3^{-*}
+ \bar{V}_{1,2,-3}^{(3)} a_2^{+} a_3^{-*}+\bar{V}_{1,-3,2}^{(3)} a_2^{-} a_3^{+*}] \delta_{1,2-3}\big\},
\label{eq:3wo}
\end{split}	
\end{equation}
\begin{equation}
\begin{split}
&	i\frac{da_1^{-}}{dt}=\omega_1^{-}a_1^{+} +\sum_{2,3}\big\{ [\bar{T}_{1,2,3}^{(1)} a_2^{+} a_3^{+}+\bar{T}_{1,2,3}^{(2)} a_2^{-} a_3^{-}
+ \bar{T}_{1,2,-3}^{(3)} a_2^{+} a_3^{-}] \delta_{1,2+3}+
[\bar{T}_{1,-2,-3}^{(1)} a_2^{+*} a_3^{+*}+ \bar{T}_{1,-2,-3}^{(2)} a_2^{-*} a_3^{-*} +
\\&
+ \bar{T}_{1,-2,3}^{(3)} a_2^{+*} a_3^{-*}]\delta_{1+2+3,0}+
[2\bar{T}_{1,2,-3}^{(1)} a_2^{+} a_3^{+*}+2\bar{T}_{1,2,-3}^{(2)} a_2^{-} a_3^{-*}
+ \bar{T}_{1,2,3}^{(3)} a_2^{+} a_3^{-*}+\bar{T}_{1,-3,-2}^{(3)} a_2^{-} a_3^{+*}] \delta_{1,2-3}\big\}.
\label{eq:3wa}
\end{split}	
\end{equation}
\rule{\textwidth}{0.4pt}
\twocolumngrid
The value of the coefficients is reported in the appendix.
These equations account for all sort of interactions between the optical and acoustical branches; however,
the large time behaviour of the system can be described by a subset of these interactions, as outlined in the next section.

\section{A statistical description: the Coupled Wave Kinetic Equations }
\label{sec:stat}
\begin{figure}[!htbp]
\includegraphics[width=0.8\columnwidth]{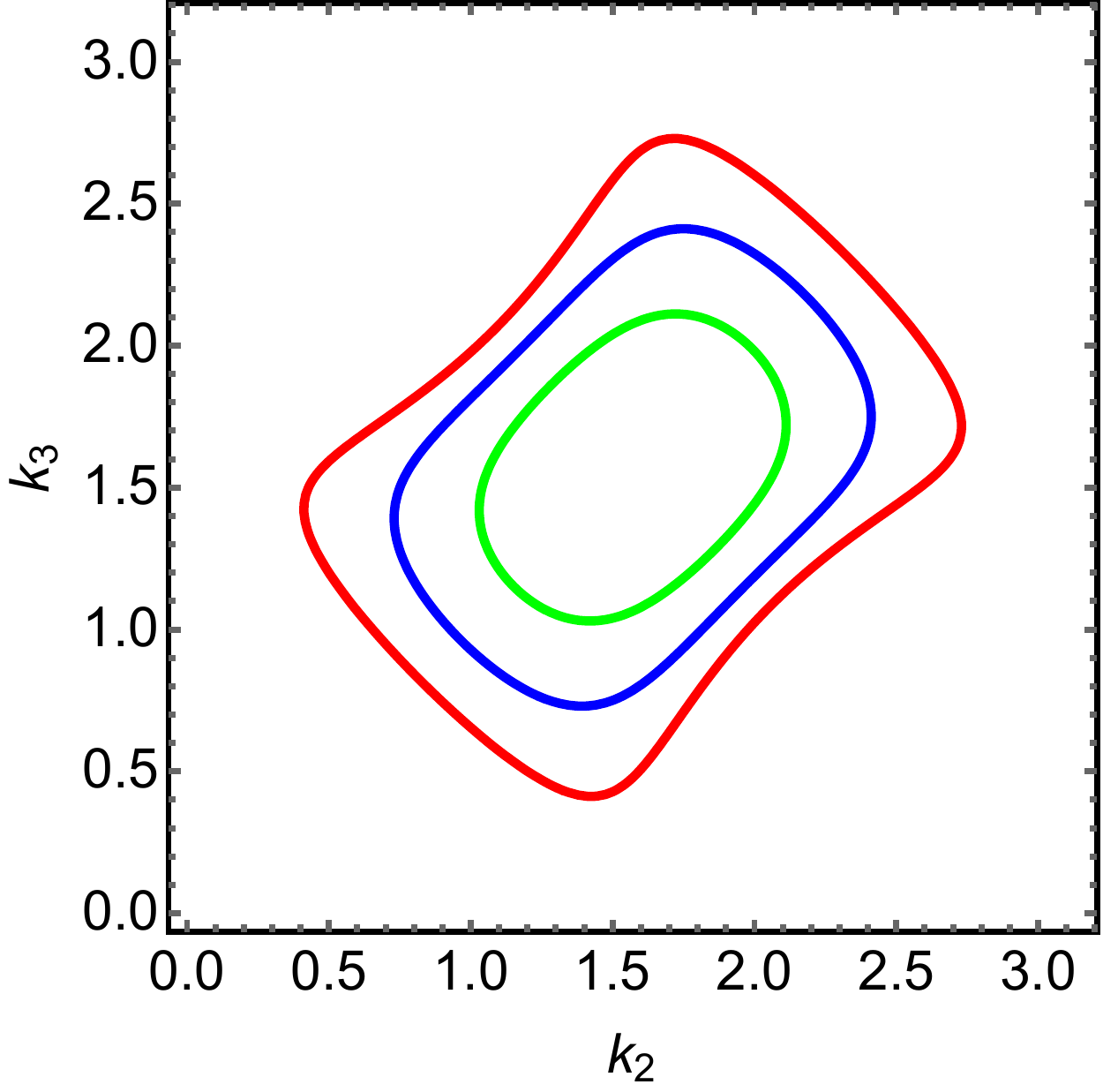}
\caption{Resonant manifold for interaction of the type $k_1=k_2+k_3$ and
 $\omega_1^+ = \omega_2^- + \omega_3^-$. The $M=1.5 m$ (red curve), $M=2m$ (blue curve) and $M=2.5m$ (green curve). The resonant manifold shrinks to a point as $M$ tends to 3 $m$ and it is empty for values of $M>3m$.}
\label{fig:figure0}
\end{figure}

The Wave Kinetic equation theory is based on the concept of resonant interactions \cite{zakharov2012kolmogorov,nazarenko2011wave}: an irreversible transfer of energy is achieved only if the resonant conditions are satisfied, which, for a three-wave interaction system, corresponds to the existence of solutions of the equations of the form:
\begin{equation}
\label{resonance1}
\begin{cases}
k_1\pm k_2\pm k_3=0 \\
\omega_1^{\pm} \pm \omega_2^{\pm} \pm \omega_3^{\pm}=0.
\end{cases}\qquad
\end{equation}
As it will be discussed later, the wave kinetic approach is obtained in the limit of large box, i.e. in the limit of $N\rightarrow \infty$ in such a way that the Fourier space becomes continuous (the discreteness in physical space is preserved). Therefore, in such a limit, wave numbers are not integers anymore and are defined in the $[0,\pi/a]$ interval.  Among all  interactions, the only possibile ones  are the following:
\begin{equation}
\begin{split}
&k_1=k_2+k_3  \\
&\omega_1^+ = \omega_2^- + \omega_3^-
\end{split}
\end{equation}
\begin{equation}
\begin{split}
&k_1=k_2-k_3 \\
&\omega_1^- = \omega_2^+ - \omega_3^-
\end{split}
\end{equation}
which are  possible only if $2\omega_{\max}^- \ge \omega_{\max}^+$, i.e. $m<M\le3m$. The resonant manifold can be easily computed numerically,
and it is shown in Figure \ref{fig:figure0} for  $m=1$ and for different values of $M=1.5,\;2,\;2.5$. As it is clear from the plot, the manifold shrinks to a single point as $M$ approaches $3$.
Whereas non-resonant terms are relevant only in the short time dynamics, we are  interested in the long time one, where resonant terms may lead to some statistically stationary state. For this reason, we disregard all the non resonant terms (formally this can be done in the weakly nonlinear regime using a \emph{near identity transformation}, \cite{zakharov2012kolmogorov}), so that the equations in the large box limit become:
\begin{equation}
\begin{split}
\label{longterm}
&i\frac{da_1^{+}}{dt}=\omega_1^+a_1^{+} + \int_0^{\pi/a} \bar{V}_{1,2,3}^{(2)}  a_2^{-} a_3^{-}
\delta_{1,2+3} dk_{2,3} \\
&i\frac{da_1^-}{dt}=\omega_1^-a_1^{-} + \int_0^{\pi/a} \bar{T}_{1,2,3}^{(3)}  a_2^{+} a_3^{-*}
\delta_{1,2-3}dk_{2,3} .
\end{split}
\end{equation}
We now assume that the system is composed by a large number of waves that are interacting through equation (\ref{longterm}). We are then interested in the in the evolution equation for the correlators $\langle a_1^{s*} a_2^{s} \rangle$ where $\langle...\rangle$ implies an ensemble average over initial random phases. Assuming statistical homogeneity of the wave field then:
\begin{equation}
\label{correlators}
\left\langle a_1^{s*}a_2^{s} \right\rangle = n_1^{s}\delta_{1,2}
\end{equation}
where $n_1^{s}=n_{k_1}^{s}$ are the wave \emph{action spectral densities} and now the $\delta_{1,2}=\delta({k_1-k_2}) $ is a Dirac Delta.
A sketch of the derivation of the kinetic equation, which does not pretend to be rigorous from a mathematical point of view, is reported in appendix; the final result is the following:
\begin{equation}
\begin{split}
	\label{WKEs}
&	\frac{\partial n_1^+}{\partial t}=4\int_{0}^{\pi/a} \vert \bar{V}_{1,2,3}^{(2)}   \vert^2
	n_1^+ n_2^- n_3^- \times\\
&	\Bigl( \frac{1}{n_1^+}-\frac{1}{n_2^-} -\frac{1}{n_3^-}\Bigr)
	\delta_{1,2+3} \delta_{\omega_1^+,\omega_2^- + \omega_3^-}dk_2dk_3, \\
&	\frac{\partial n_1^-}{\partial t}=8\int_{0}^{\pi/a}  \vert \bar{V}_{2,1,3}^{(2)}  \vert^2
	n_1^- n_2^+ n_3^- \times\\&
	\Bigl( \frac{1}{n_1^-}-\frac{1}{n_2^+} +\frac{1}{n_3^-}\Bigr)
	\delta_{1,2-3} \delta_{\omega_1^-, \omega_2^+ - \omega_3^-}dk_2dk_3,
\end{split}
\end{equation}
i.e. two coupled equations for the evolution of the wave action spectral density of the optical and acoustic modes. We can observe that, because of the presence of the two $\delta$s in the right hand side, the integral is not zero only if resonance conditions are satisfied, otherwise the spectral density does not evolve in time as in the linear case.
\subsection{\label{sec:ConsevedQ} Collision invariants, $H$- theorem and thermodynamic solution}

The integrals in the right hand side of equations in (\ref{WKEs}) can be seen as \emph{collision integrals} of the type in the celebrated  Boltzmann equation for a gas of interacting particles. It is not difficult to verify that the \emph{total energy},
\begin{equation}
\label{total_energy}
E=\int_{0}^{\pi/a} (\omega_k^+n_k^+ + \omega_k^- n_k^-) =
\int_{0}^{\pi/a} (\mathcal{E}_k^+ + \mathcal{E}_k^-) dk,
\end{equation}
is a conserved quantity, where $\mathcal{E}_k^\pm$ are energy densities for the optical and acoustical modes.
Moreover,
if we define an \emph{entropy} as
\begin{equation}
\label{entropy}
	S=\int_0^{\pi/a} \ln [ n_k^+ n_k^-]dk,
\end{equation}
an \emph{H-theorem} can be proved, i.e. $dS/dt\ge0$.
At the thermodynamic equilibrium $dS/dt=0$ and we get the stationary solutions of (\ref{WKEs}) at equilibrium, i.e. the \emph{Rayleigh-Jeans distributions}:
\begin{equation}
\label{RJ}
\mathcal{E}_k^\pm=\omega_k^\pm n_k^\pm= T.
\end{equation}
Combining~\eqref{total_energy} and~\eqref{RJ}, we obtain
\begin{equation}
E=2\pi T/a.
\end{equation}
This implies that, as expected, the equilibrium is characterized by the \emph{equipartition of energy} among all the degrees of freedom, i.e., the Fourier modes associated to the diagonalized variables. Note that the $n_k^\pm$ are variables that have been obtained through a number of transformations. It becomes then important to go back to the original variables and characterize the equilibrium in terms of them.
Inverting equations in (\ref{canonical_transform}) and computing the modulus square, we get:
\begin{equation}
\label{Q_and_R_MOD}
\begin{split}
&\lvert Q_k \rvert^2=
\lvert \widetilde{Q}_k^{+}  \rvert^2 + \lvert \widetilde{Q}_k^{-}  \rvert^2 + \widetilde{Q}_k^{+} \widetilde{Q}_k^{-*} + \widetilde{Q}_k^{+*}\widetilde{Q}_k^{-} \\
&\lvert R_k \rvert^2=
(\beta_k^+)^2\lvert \widetilde{Q}_k^+  \rvert^2 + (\beta_k^-)^2\lvert \widetilde{Q}_k^{-}  \rvert^2 - \frac{m}{M}
(\widetilde{Q}_k^+ \widetilde{Q}_k^{-*} + \widetilde{Q}_k^{+*} \widetilde{Q}_k^{-}).
\end{split}
\end{equation}
Using equation (\ref{wave-action1}) to express $\widetilde{Q}_k^{\pm}$ in terms of
the normal variables $a_k^{\pm}$, taking the expectation value (with random phase approximation) and finally substituting the equilibrium solution, equation (\ref{RJ}), we get:
\begin{equation}
\label{spectra_diagonal}
\begin{split}
&\langle \lvert \widetilde{Q}_k^{\pm}  \rvert^2 \rangle =
\frac{T}{\mu_k^{\pm} (\omega_k^{\pm})^2},\;\;\;\;\; \langle \widetilde{Q}_k^{\pm} \widetilde{Q}_k^{\mp*} \rangle  = 0.
\end{split}
\end{equation}
Taking the expectation value of (\ref{Q_and_R_MOD}) and inserting (\ref{spectra_diagonal}), we obtain
\begin{equation}
\label{spectra}
\langle \lvert Q_k \rvert^2 \rangle = \langle \lvert R_k \rvert^2 \rangle
=\frac{T}{2\chi} \csc^2(a k);
\end{equation}
in a similar way, we also obtain
\begin{equation}
\label{mixed_mean_values}
\langle Q_kR_{k}^{*} \rangle =\langle Q_k^*R_{k} \rangle=
\frac{T}{2\chi} \cot(a k)\csc(a k).
\end{equation}
Proceeding as before, we obtain
\begin{equation}
\label{P_G_mean}
\langle\lvert\widetilde{P}_k^{\pm}\rvert^2 \rangle =  \mu_k^{\pm}  T,
\end{equation}
and
\begin{equation}
\langle\lvert P_k\rvert^2\rangle=mT, \quad \,\,
\langle\lvert G_k\rvert^2\rangle=MT.
\label{eq:kinetic}
\end{equation}

\section{Numerical simulations and verification of the theoretical predictions}\label{sec:num}
\begin{figure}[ht]
\begin{minipage}[b]{0.49\linewidth}
\centering
\includegraphics[width=1\columnwidth]{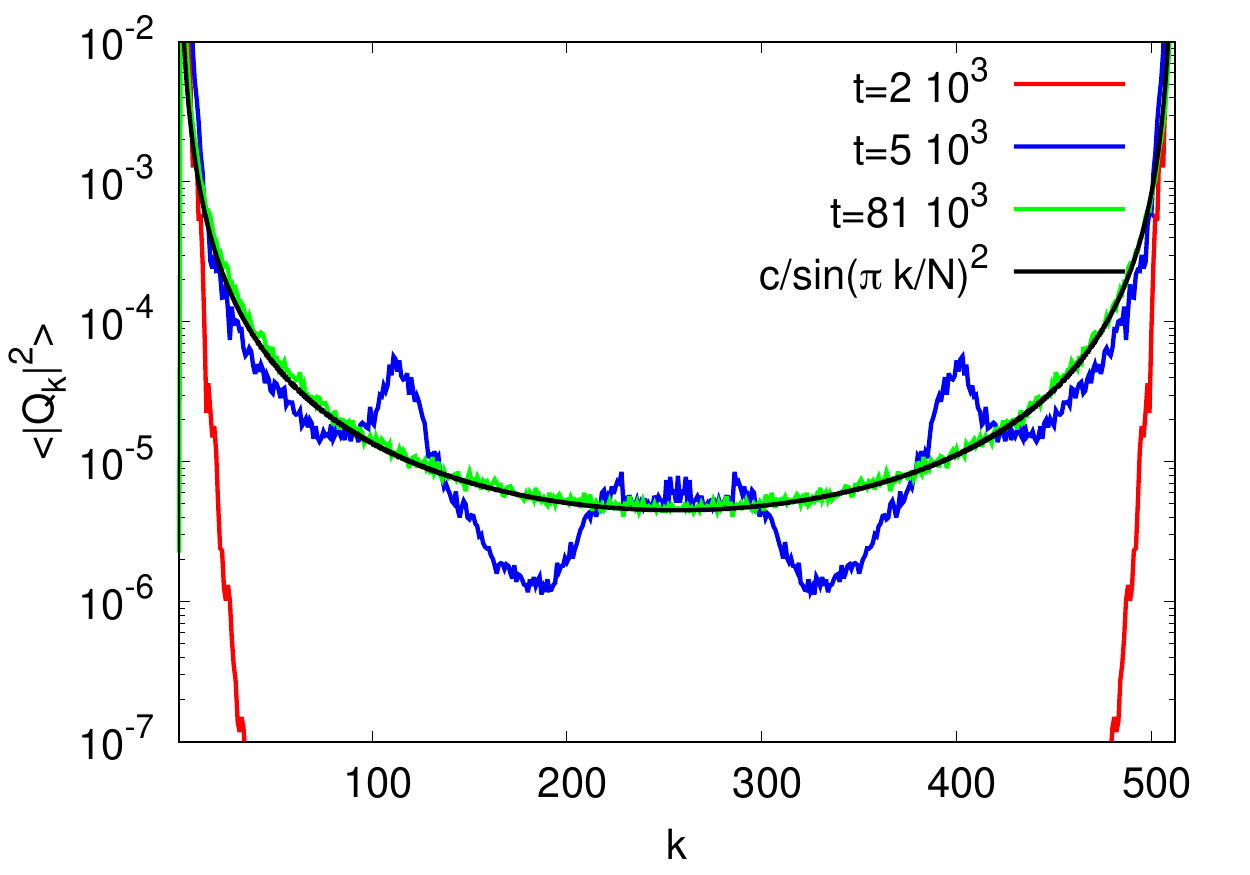}
\end{minipage}
\begin{minipage}[b]{0.49\linewidth}
\centering
\includegraphics[width=1\columnwidth]{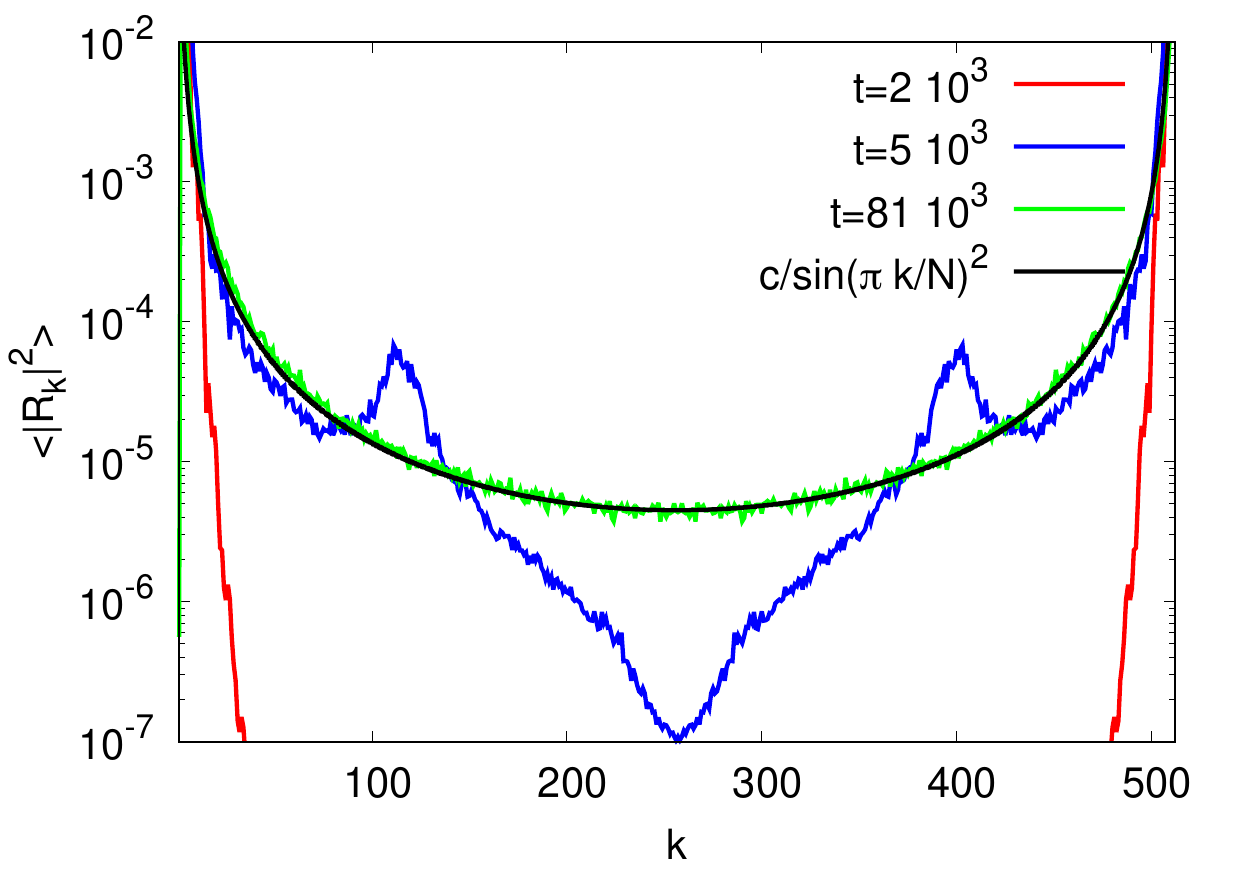}
\end{minipage}
\caption{$\langle|Q_k|^2\rangle)$ (left) and $\langle|R_k|^2\rangle)$ (right) as a function of wave number for different time. $\langle|Q_k|^2\rangle)$ and $\langle|R_k|^2\rangle)$ represent the expectation value of the modulus square of the Fourier amplitudes of the position of the masses $M$ and $m$, respectivelly. The black line corresponds to the theoretical prediction in equation (\ref{spectra}) with constant $c=T/(2\chi)=4.5\times10^{-6}$. Initial conditions are provided by equation (\ref{initial}) with $M=2m$.}
\label{fig:figure1}
\end{figure}

\begin{figure}[ht]
\begin{minipage}[b]{0.49\linewidth}
\centering
\includegraphics[width=1.\columnwidth]{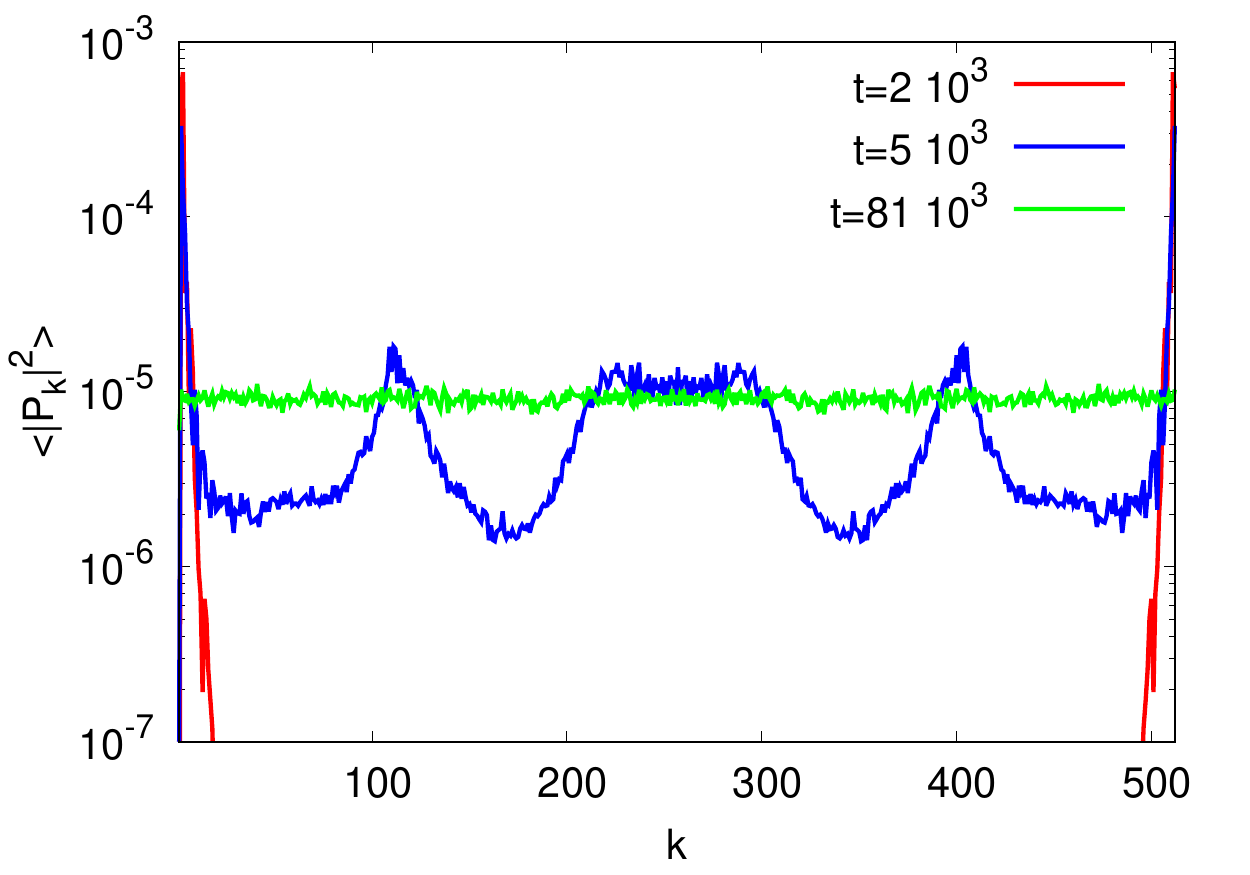}
\end{minipage}
\begin{minipage}[b]{0.49\linewidth}
\centering
\includegraphics[width=1.\columnwidth]{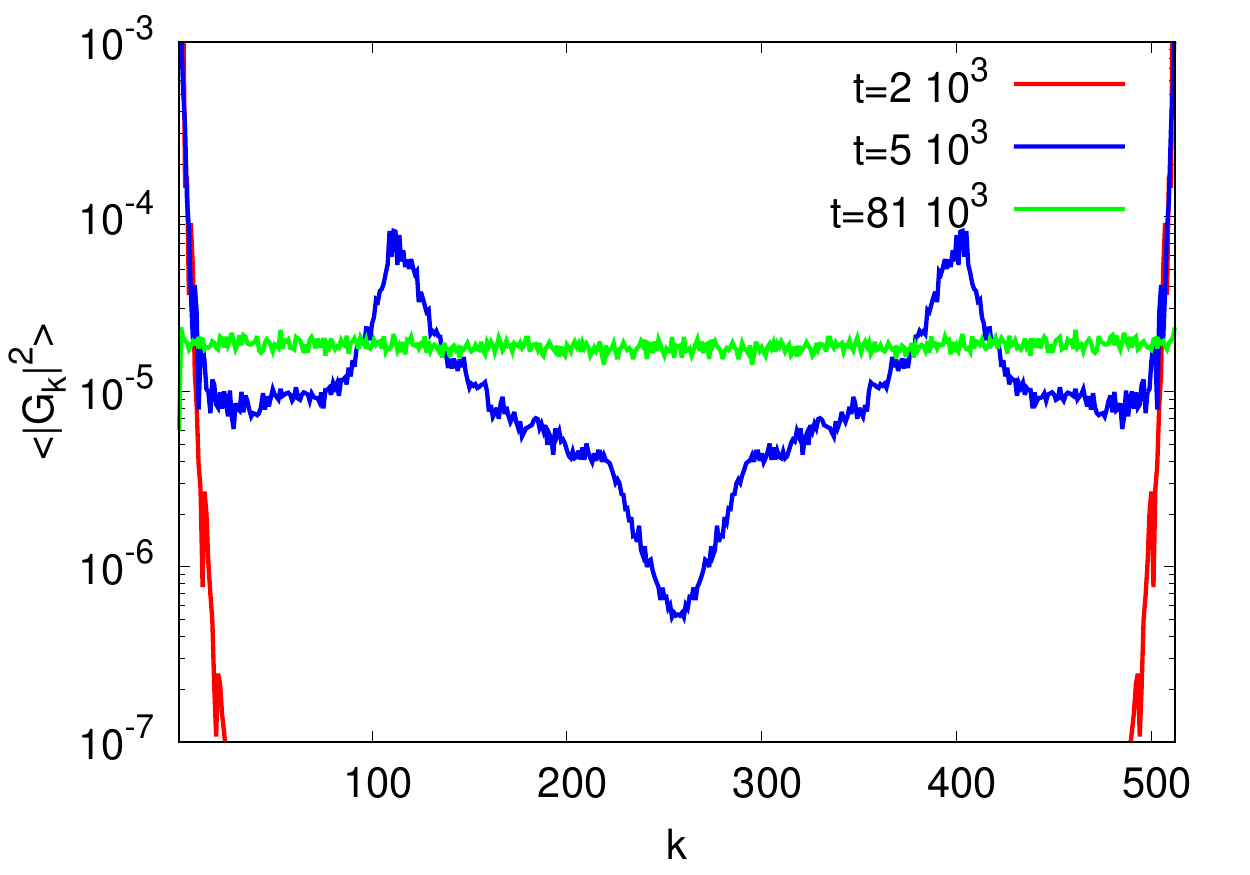}
\end{minipage}
\caption{Expectation value for the kinetic energy density $\langle|P_k|^2\rangle)$ (left) and $\langle|G_k|^2\rangle)$ (right) associated with masses $m$ and $M$, respectively,  at different time steps. Initial conditions are provided by equation (\ref{initial}) with $M=2m$.}
\label{fig:figure2}
\end{figure}

\begin{figure}[!htbp]
\includegraphics[width=0.9\columnwidth]{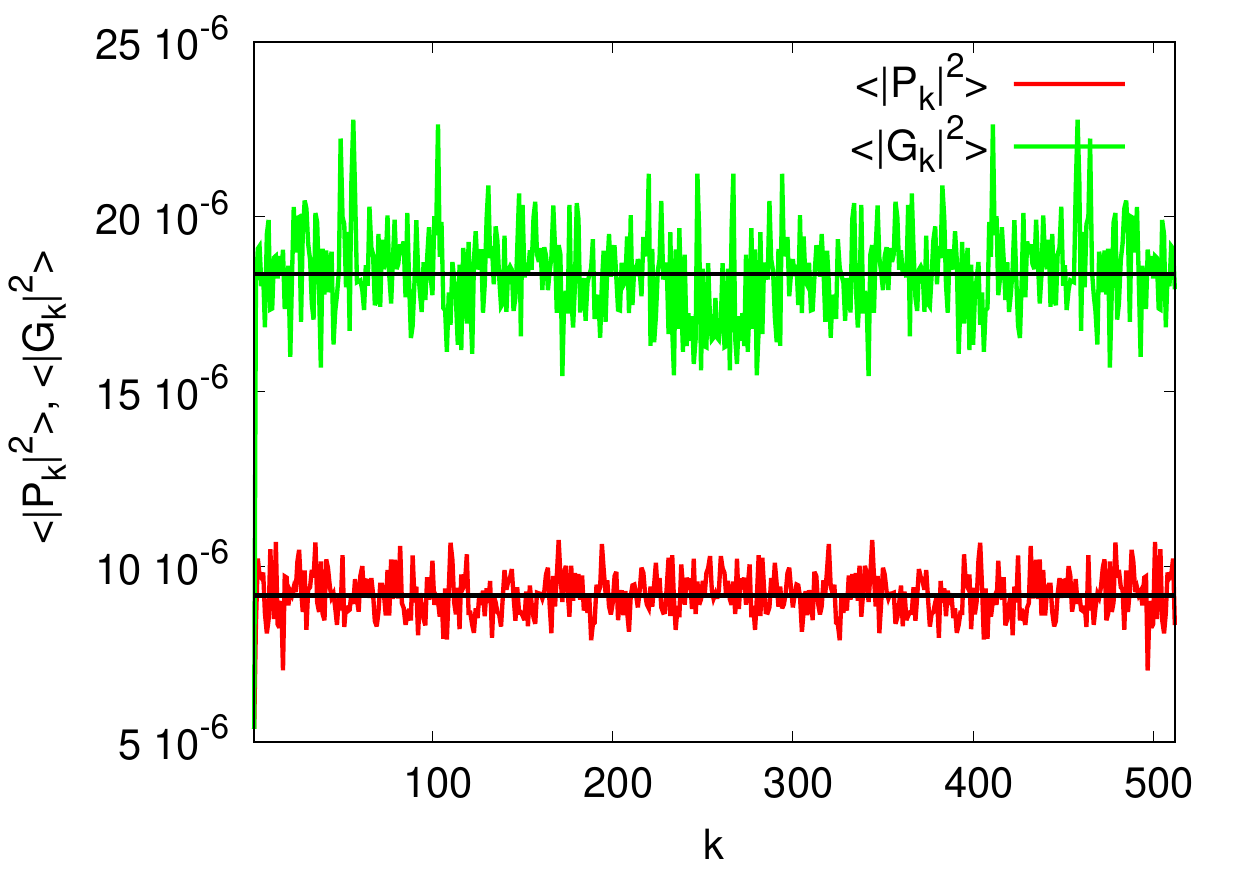}
\caption{$\langle|P_k|^2\rangle)$ and $\langle|G_k|^2\rangle)$ at time $t= 10^4$. The black horizontal lines
represents the mean  value in $k$ for each density, $1.836 \time 10^{-5}$ and $0.918 \time 10^{-6}$. Their ratio is 2, as predicted by equations in (\ref{eq:kinetic}).   }
\label{fig:figure5}
\end{figure}

The theoretical predictions discussed in the above section are now  compared with long time simulations of the  deterministic equations of motion. We have developed a numerical code for solving the equations in
(\ref{eqmotionnl1}) and  (\ref{eqmotionn2}) using a 4-th order Runge-Kutta method with periodic boundary conditions. We have verified that in all our simulations the Hamiltonian is preserved with a relative error of less than $1\%$. Our simulations are performed in the same spirit as the one of Fermi and collaborators \cite{fermi1955alamos}; here, initial data are provided by the sum of two long sinusoidal waves:
 \begin{equation}
y_j(t=0)=A \left[\sin\left(\frac{ \pi j}{ N}\right)+ \sin\left(\frac{2 \pi j}{ N}+\phi\right)\right]
\label{initial}
 \end{equation}
 and $\dot y_j(t=0)=0$ with $j=1,2,...,2N$. We have introduced a phase $\phi$ and we have run  200 simulations,
 each with a different random phase distributed in the interval $[0,2\pi]$.
 Observables are obtained by performing ensemble averages over all the members of the ensemble.
 The parameter $\alpha$ in front of the nonlinear terms is set to 1 and the degree of the nonlinearity in
 the simulation is ruled by the amplitude $A$ in (\ref{initial}). In Figure \ref{fig:figure1} we show three snapshots
 of $\langle|Q_k|^2\rangle)$ and $\langle|R_k|^2\rangle)$ at different times of a simulation characterized by $A=10$, $M=2$ and $m=1$.
For large times the systems reaches  its thermal equilibrium, see green curve in the figure;
the theoretical prediction is also plotted, displaying an excellent agreement with numerics.
While for large times, the equilibrium for the observables $\langle|Q_k|^2\rangle$ and $\langle|R_k|^2\rangle$
is proportional to $\csc(\pi k/N)^2$, for the spectral kinetic energy densities, $\langle|P_k|^2\rangle$ and $\langle|G_k|^2\rangle$, the predictions correspond to an equipartition among the Fourier modes, see
equation (\ref{eq:kinetic}). Figure \ref{fig:figure2} shows the spectral kinetic energy density associated with masses $M$ and $m$, respectively, as a function of wavenumber $k$ for different instant of time. The simulations show that the large time behavior  is characterized by a constant kinetic energy  density. Interestingly, the theory predicts that the ratio between  $\langle|P_k|^2\rangle$ and $\langle|G_k|^2\rangle$ should corresponds to the ratio of the masses (2 in the present case). This is displayed clearly in Figure  \ref{fig:figure5}, where the $\langle|P_k|^2\rangle$ and $\langle|G_k|^2\rangle$ are represented in the same plot, once equilibrium has been reached. The ratio between the mean value in $k$ of the two curves is 2, as predicted.
Similar results (not shown here)  can be obtained for different values of the initial amplitude $A$ or different ratio of masses, but always larger than 1 and lesser or equal to 3.

\begin{figure}[ht]
\begin{minipage}[b]{0.49\linewidth}
\centering
\includegraphics[width=1.\columnwidth]{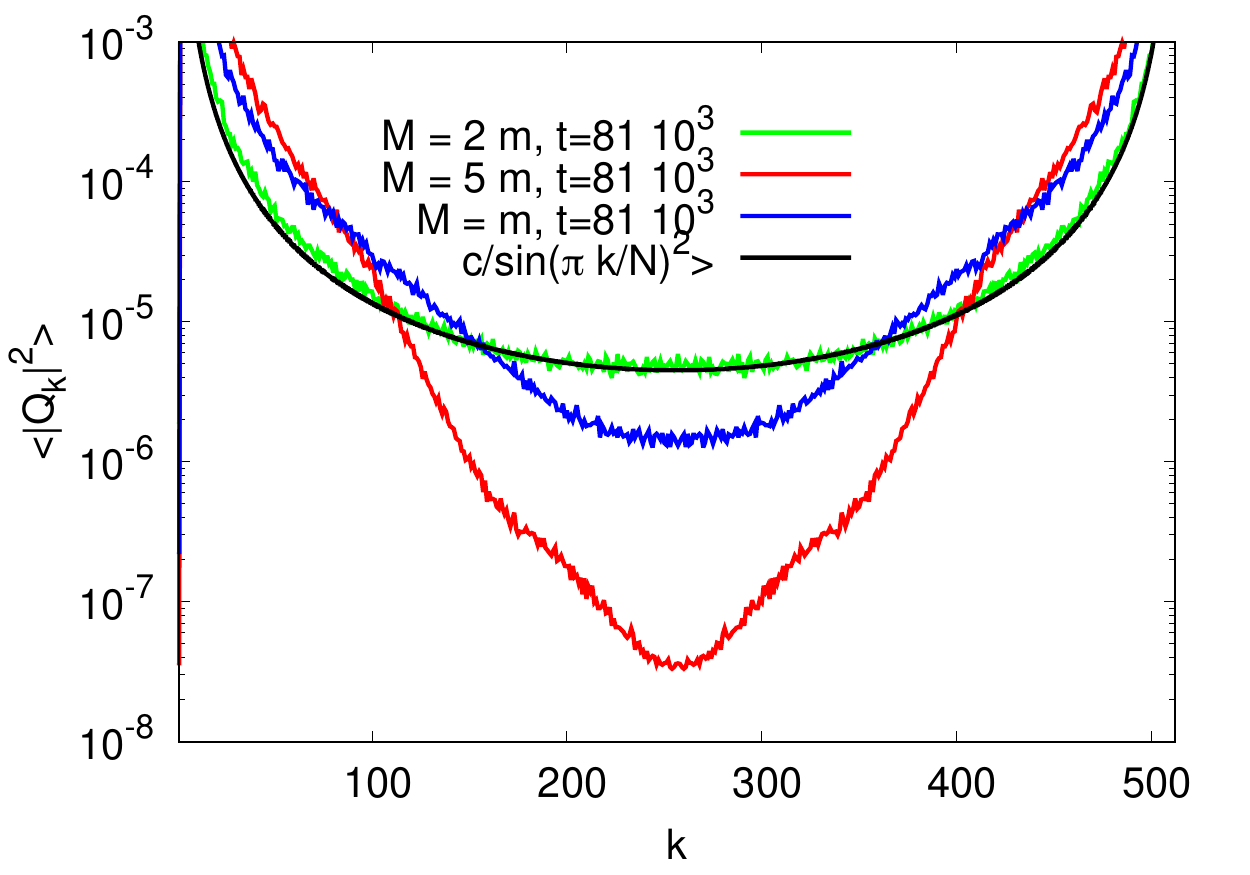}
\end{minipage}
\begin{minipage}[b]{0.49\linewidth}
\centering
\includegraphics[width=1.\columnwidth]{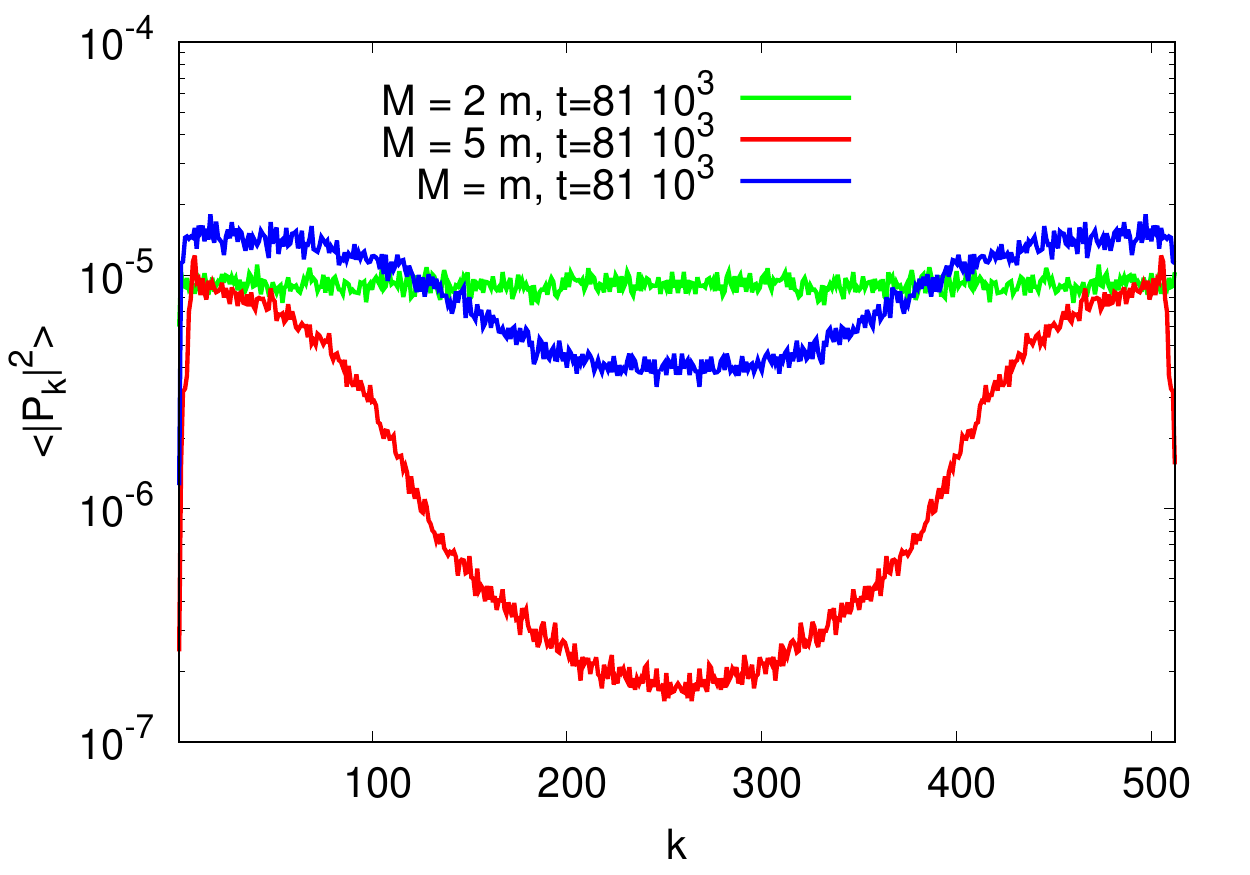}
\end{minipage}
\caption{ $\langle|Q_k|^2\rangle$ (left) and $\langle|P_k|^2\rangle$ (right) as a function of $k$  at time  $t= 81\times 10^3$ for  $M= 2m$ (green line), $M=m$ in (blue line) and $ M=5 m$  in (red line).}
\label{fig:figure7}
\end{figure}

With respect to the standard $\alpha$-FPUT model where all masses are equal, we predict that the relaxation time is much faster; the reason for such statement relies on the fact that the evolution in time of the wave action spectral density function is described by a three-wave system and not by a four-wave system. The presence of two branches in the dispersion relation allows for exact three-wave resonant interactions. Moreover, it was found that resonant interactions are possibile only if the ratio between heavy and light masses is less or equal to 3. We now use numerical simulations to test such theoretical findings: we perform the same simulations as those previously described but for $m=M$ and $M=5m$. The results are displayed in Figure \ref{fig:figure7}, where the $\langle|Q_k|^2\rangle$ and $\langle|P_k|^2\rangle$  are shown as a function of $k$ at fixed time for different mass ratios. The Figures highlight the fact that, as expected, the fastest evolution that reaches first the thermalized state is characterized by $M=2m$. Exact three-wave resonant interactions in case of $M=5m$ and $M=m$
do not exists and the evolution of the spectra is related to either quasi-resonant three-wave interactions related to the finite nonlinearity effect  or higher order interactions.
 \section{\label{Conclusions} CONCLUSIONS}
Since the pioneering work by E. Fermi and collaborators  \cite{fermi1955alamos}, a lot of theoretical and numerical work  has been done in the study of thermalization of one dimensional chains (see the latest review dated already 2008 \cite{gallavotti2007fermi}). The $\alpha$- and $\beta$-FPUT systems have been widely studied in different contexts and, nowadays, we know from numerical simulations that their long time behavior is characterized by an equilibrium which is very close to equipartition of energy among the Fourier modes \cite{benettin2011time,Ponno2011}. This statement is not based on any rigorous theory but rather on long and robust numerical simulations. The estimation of the thermalization time scale in the limit of small nonlinearity can be obtained using the wave turbulence approach, which, despite being not mathematically rigorous, is based on a solid physical background. The time scale for thermalization in the above systems is much longer than time scale of the dynamical equation. The reason is that the mechanism of irreversible transfer of energy between modes is the resonant interaction among waves: because of the shape of the dispersion relation, both the monoatomic $\alpha$- and $\beta$-FPUT are characterized, in the large box limit, by an energy transfer ruled by four-wave resonant interactions.

 In this paper we have analyzed the diatomic $\alpha$-FPUT chain, i.e. a system of alternating masses with a cubic potential. Interestingly, the introduction of such interchanging between two different masses in the chain has considerable effects on the thermalization time scale: the dispersion relation drastically changes and two branches, the optical and the acoustical, appear. This system allows for three-wave resonant interactions but only between two acoustic and one optical wave (no exchange between waves of the same branch is possible or between two optical waves and one acoustic wave). Moreover, it turns out that resonances take place only for masses such that the ratio between heavy and light particles is smaller or equal to 3 and greater than 1. Under such constraints, two coupled wave kinetic equations, each describing the evolution in time of the wave action spectral density function of the acoustical and optical modes, can be formally derived. We show that it is possible to introduce an entropy for which an $H$- theorem holds; this implies that an irreversible dynamics towards an equilibrium solution which corresponds to equipartition of energy (in the diagonalized variables) takes place. The equilibrium solution is found in the diagonalized variables and then, inverting the canonical transformations, it can be written in terms of the  original variables (in Fourier space) of the system. Numerical computation of the microscopic dynamics are in very good agreement with the theoretical predictions; moreover, a number of simulations have also been performed in order to show that the relaxation time scale for the diatomic case for $m<M\le3m$ is smaller with respect to the monatomic case characterized by $m=M$ or the case  with $M=5m$. The present results highlight once more the power of the wave turbulence approach for studying the statistical properties of nonlinear dispersive waves in the limit of small nonlinearity.


\begin{acknowledgments}
G.D. acknowledges the support from the Australian Research Council Discovery Project DP 190101190.
M.O. was supported by the ``Departments of Excellence 2018-2022'' Grant awarded by the Italian Ministry of
Education, University and Research (MIUR) (L.232/2016). M.O. was supported by Simons Collaboration on
Wave Turbulence, Grant No. 617006 and by the European Commission H2020 FET Open
24 ``Boheme'' grant no. 863179.
\end{acknowledgments}
\bibliography{references}
\onecolumngrid
\appendix

\section{Diagonalization }
The  equations of motion in the Fourier space take the following form:
\begin{subequations}
	\label{eqmotionfourier}
	\small{
	\begin{eqnarray}
	&\ddot{Q}_1&=\frac{2\chi}{m} [R_1 \cos(aq_1)-Q_1]+\frac{2i}{\sqrt{N}} \frac{\alpha}{m}
	\sum_{2,3}\bigl\{(-1)^l R_2R_3\sin(aq_1)-2R_2Q_3\sin(aq_2)\bigr\} \delta_{1,2+3}
	\label{eqmotionfourier1}\\
	&\ddot{R}_1&=\frac{2\chi}{M} [Q_1 \cos(aq_1)-R_1]+\frac{2i}{\sqrt{N}} \frac{\alpha}{M}
	\sum_{2,3}\bigl\{ Q_2Q_3 \sin(aq_1)-2(-1)^l Q_2R_3\sin(aq_2)\bigr\} \delta_{1,2+3}.
	\label{eqmotionfourier2}
	\end{eqnarray}}
\end{subequations}
Equation (\ref{eqmotionfourier}) are coupled, even in their linear part. To diagonalize the system,  we first write the linear part in matrix form:
\begin{equation}
\label{linearmatrix}
\ddot{\vec{Q}}=A\vec{Q},
\end{equation}
where
\begin{equation}
\vec{Q}=
\begin{bmatrix}
Q_k \\
R_k
\end{bmatrix}
\quad
\mathrm{and}
\quad
A=
\begin{bmatrix}
-\frac{2\chi}{m} & \frac{2\chi}{m}\cos(aq_k) \\
\frac{2\chi}{M}\cos(aq_k) & -\frac{2\chi}{M}.
\end{bmatrix}
\end{equation}
Solving the secular equation, we find the eigenvalues $\lambda_\pm$ and thus the diagonal matrix $A_D$:
\begin{equation}
\label{diagonalmatrix}
A_D=
\begin{bmatrix}
-\omega_{+}^2(q_k) & 0 \\
0 & -\omega_{-}^2(q_k)
\end{bmatrix}
.
\end{equation}
Solving $A \vec{u}_\pm=\lambda_\pm\vec{u}_\pm$, we get two eigenvectors which constitute the change-of -basis matrix:
\begin{equation}
\label{Pmatrix}
X=
\begin{bmatrix}
1 & 1  \\
\beta_k^+ & \beta_k^-
\end{bmatrix}, \quad\quad
X^{-1}=
\begin{bmatrix}
\frac{m}{\mu_k^+} & \frac{\beta_k^+ M}{\mu_k^+}  \\
\frac{m}{\mu_k^-} & \frac{\beta_k^- M}{\mu_k^-},
\end{bmatrix},
\end{equation}
where $\beta_k^\pm$ and $\mu_k^\pm$ are defined in~\eqref{betarelation} and~\eqref{e:mu_k}, respectively.
Noting that
\begin{equation}
A = XA_D X^{-1},
\end{equation}
we can write~\eqref{linearmatrix} as
\begin{equation}
\label{diagonalization}
\ddot{\vec{\widetilde{Q}}}=A_D\vec{\widetilde{Q}}
\end{equation}
where
\begin{equation}
\label{diag_linearsystem}
\vec{\widetilde{Q}}=X^{-1}\vec{Q}
\equiv\begin{bmatrix}
\widetilde{Q}_k^+\\
\widetilde{Q}_k^-
\end{bmatrix}.
\end{equation}
The Hamiltonian  in the new variables takes the following form:
\small{
\begin{eqnarray}
\label{hamiltontransf}
H&=&\sum_{k} \biggl\{
\frac{\lvert \widetilde{P}_k^+ \rvert^2}{2 \mu_k^+} +
\frac{\lvert \widetilde{P}_k^- \rvert^2}{2\mu_k^-} +
\frac{1}{2} \mu_k^+ \omega_+^2(q_k) \lvert \widetilde{Q}_k^+ \rvert^2+
\frac{1}{2} \mu_k^- \omega_-^2(q_k) \lvert \widetilde{Q}_k^- \rvert^2
\biggr\} + \\
&+&2i \alpha \sum_{1,2,3} \bigl\{
W_{1,2,3}^{(1)} \widetilde{Q}^+_1\widetilde{Q}^+_2\widetilde{Q}^+_3 +
W_{1,2,3}^{(2)} \widetilde{Q}^-_1\widetilde{Q}^-_2\widetilde{Q}^-_3 +
W_{{1,2,3}}^{(3)} \widetilde{Q}^+_1\widetilde{Q}^-_2\widetilde{Q}^-_3 +
W_{1,2,3}^{(4)} \widetilde{Q}^-_1\widetilde{Q}^+_2\widetilde{Q}^+_3
\bigr\}\delta_{k_1+k_2+k_3,0} \notag
\end{eqnarray}}
\normalsize
with
\begin{subequations}
	\label{3.86}
	\begin{align}
	W_{1,2,3}^{(1)} &= [A_{1,2,3}^{+++})+A_{2,1,3}^{+++}+A_{3,2,1}^{+++}]/3
	\label{3.86a} \\
	W_{1,2,3}^{(2)} &=  [A_{1,2,3}^{---}+A_{2,1,3}^{---}+A_{3,2,1}^{---}]/3
	\label{3.86b} \\
	W_{1,2,3}^{(3)} &=A_{1,2,3}^{+--} + A_{2,1,3}^{-+-} + A_{3,1,2}^{-+-}
    \label{3.86c}\\
	W_{1,2,3}^{(4)} &= A_{1,2,3}^{-++} +A_{2,1,3}^{+-+} + A_{3,1,2}^{+-+},
	\label{3.86d}
	\end{align}
\end{subequations}
and
\begin{equation}
A_{1,2,3}^{s_1,s_2,s_3}=(\beta_1^{s_1} + (-1)^l \beta_2^{s_2} \beta_3^{s_3} )\sin(aq_1).
\end{equation}

Finally the equations of motion for the optical and acoustic branches can be written as:
\begin{subequations}
	\label{nlsystemdiagonal}
	\begin{eqnarray}
	\ddot{\widetilde{Q}}^+_1+(\omega_1^+)^2\widetilde{Q}^+_1&=&
	\frac{2i\alpha}{\mu_1^+}\sum_{2,3}
	\bigl\{V_{1,2,3}^{(1)}\widetilde{Q}^+_2\widetilde{Q}^+_3+
	V_{1,2,3}^{(2)}\widetilde{Q}^-_2\widetilde{Q}^-_3+
	V_{1,2,3}^{(3)}\widetilde{Q}^+_2\widetilde{Q}^-_3
		\bigr\} \delta_{1,2+3} \label{nlsystemdiagonal1}\\ 
	\ddot{\widetilde{Q}}^-_1+(\omega_1^-)^2\widetilde{Q}^-_1&=&
	\frac{2i \alpha}{\mu_1^-}\sum_{2,3}
	\bigl\{T_{1,2,3}^{(1)}\widetilde{Q}^+_2\widetilde{Q}^+_3+
		T_{1,2,3}^{(2)}\widetilde{Q}^-_2\widetilde{Q}^-_3+
	T_{1,2,-3}^{(3)}\widetilde{Q}^+_2\widetilde{Q}^-_3
	\bigr\} \delta_{1,2+3}\;\;\;,
	 \label{nlsystemdiagonal2}
	\end{eqnarray}
\end{subequations}
with
\begin{equation}
\begin{split}
\label{3.90}
 & V_{1,2,3}^{(1)}=-3 W_{-1,2,3}^{(1)}\\
 &V_{1,2,3}^{(2)}=- W_{-1,2,3}^{(3)}\\
& V_{1,2,3}^{(3)}= -2W_{3,-1,2}^{(4)}
 \end{split}
\end{equation}
and
\begin{subequations}
	\label{3.95}
\begin{align}
T_{1,2,3}^{(1)}&=-W_{-1,2,3}^{(4)} \label{3.95a}\\
T_{1,2,3}^{(2)}&=-3W_{-1,2,3}^{(2)}\label{3.95b}\\
T_{1,2,3}^{(3)}&=-2W_{2,-1,-3}^{(3)}\label{3.95c}
  \end{align}
\end{subequations}
The nonlinear terms account for optical-optical-optical, optical-optical-acoustic, optical-acoustic-acoustic and acoustic-acoustic-acoustic interactions.

\section{Coefficients in (\ref{eq:3wo}) and (\ref{eq:3wa}) }
\begin{equation}
\begin{split}
&\bar{V}_{1,2,3}^{(1)}=\gamma_{1,2,3}^{+ + +}V_{1,2,3}^{(1)}, \;\;\;
\bar{V}_{1,2,3}^{(2)} =\gamma_{1,2,3}^{+ --}V_{1,2,3}^{(2)}, \;\;\;
\bar{V}_{1,2,3}^{(3)} =\gamma_{1,2,3}^{+ +-}V_{1,2,3}^{(3)}, \;\;\; \\
\end{split}
\end{equation}


\begin{equation}
\begin{split}
&\bar{T}_{1,2,3}^{(1)}=\gamma_{1,2,3}^{- + +}T_{1,2,3}^{(1)}, \;\;\;
\bar{T}_{1,2,3}^{(2)} =\gamma_{1,2,3}^{- --}T_{1,2,3}^{(2)}, \;\;\;
\bar{T}_{1,2,3}^{(3)} =\gamma_{1,2,3}^{- +-}T_{1,2,3}^{(3)}, \;\;\;
\end{split}
\end{equation}
\begin{equation}
\gamma_{1,2,3}^{s_1 s_2 s_3}=
-\frac{i \alpha}{\sqrt{2 \mu_1^{s_1}\mu_2^{s_2}\mu_3^{s_3}\omega_1^{s_1}\omega_2^{s_2}\omega_3^{s_3}}}
\end{equation}

\section{Formal Derivation of the Coupled wave kinetic equations}
Multiplying the first of (\ref{longterm}) by $a^{+*}_1$ and the complex conjugate equation by $a^{+}_1$, then subtracting the two and taking the expectation value with respect to initial data characterized by random phases and amplitudes, we get
\begin{equation}
\label{eq_2nd_order_corr}
\frac{\partial n_1^+}{\partial t}=\Im \int_0^{\pi/a}2\bar{V}_{1,2,3}^{(2)} \left\langle a^{+*}_1 a^{-}_2 a^{-}_3\right\rangle \delta_{1,2+3}
dk_2 dk_3,
\end{equation}
where $\Im$ denotes the imaginary part of the expression (note that the $\delta$ is now a Dirac Delta). By using~\eqref{longterm} we write an evolution equation of the higher order correlator in~\eqref{eq_2nd_order_corr} in the following form:
\begin{equation}
\label{3dorder_corr_eq}
\begin{split}
\Bigl[i\frac{\partial}{\partial t }+(\omega_1^+ -\omega_2^- - \omega_3^-)\Bigr]
\left\langle a^{+*}_1 a^{-}_2 a^{-}_3 \right\rangle &=
\int_0^{\pi/a}\big[\bar{T}^{(3)}_{2,4,5} \left\langle a^{+*}_1 a^{-}_3 a^{+}_4 a^{-*}_5 \right\rangle \delta_{2,4-5} + \\
+ \bar{T}_{3,4,5}^{(3)} \left\langle a^{+*}_1 a^{-}_2 a^{+}_4 a^{-*}_5 \right\rangle \delta_{3,4-5}&-
 \bar{V}_{1,4,5}^{(2)*} \left\langle a^{-}_2a^{-}_3a^{-*}_4a^{-*}_5 \right\rangle \delta_{1,4+5}\big] dk_4 dk_5.
\end{split}
\end{equation}
Because (\ref{3dorder_corr_eq}) depends on a fourth order correlator, to close the equation we have to use the \emph{Wick's selection rule}, for which a fourth-order correlator can be written as the sum of second-order correlators, so that for example
\begin{equation}
\label{4thorder_corr}
\left\langle a^{-}_2 a^{-}_3 a^{-*}_4 a^{-*}_5 \right\rangle =
\left\langle a^{-}_2 a^{-*}_4 \right\rangle \left\langle a^{-}_3 a^{-*}_5 \right\rangle +
\left\langle a^{-}_2 a^{-*}_5 \right\rangle \left\langle a^{-}_3 a^{-*}_4 \right\rangle=
n_4^-n_5^- (\delta_{4,2}\delta_{5,3}+\delta_{5,2}\delta_{4,3}).
\end{equation}
We assume that that mixed correlators are negligible  because of the assumptions of random phases, whereas
\begin{equation}
\label{corrIII}
\left\langle a^{+*}_1 a^{-}_2 a^{-}_3 \right\rangle=J_{1,2,3} \, \delta_{1,2+3}
\end{equation}
is finite and $J_{1,2,3}$ is a quantity which needs to be determined.
Applying~\eqref{4thorder_corr} and~\eqref{corrIII} to~\eqref{3dorder_corr_eq}, we obtain
\begin{equation}
\label{IIIorder_eq}
\Bigl[i\frac{\partial}{\partial t}+(\omega_1^+ -\omega_2^- - \omega_3^-)\Bigr]
J_{1,2,3} \, \delta_{1,2+3} = 2\bar{V}_{1,2,3}^{(2)*}
(n_1^+ n_2^- + n_1^+n_3^- -n_2^-n_3^-) \delta_{1,2+3},
\end{equation}
where we have used the fact that $\bar T_{2,1,3}^{(3)}=\bar T_{3,1,2}^{(3)}=2 \bar{V}_{1,2,3}^{(2)*}$.
Assuming that the spectral density function evolves in a much slower temporal scale
with respect to the correlator $J_{1,2,3}$, we can consider $n_k^\pm$ constant in first approximation. We can therefore solve~\eqref{IIIorder_eq} to obtain
\begin{equation}
\label{eq_corr3_solved}
J_{1,2,3}=Ce^{i\Delta\omega t} + \frac{2V_{1,2,3}^{(2)*}
(n_1^+ n_2^- + n_1^+n_3^- -n_2^-n_3^-)}{\Delta\omega},
\end{equation}
where $\Delta\omega=\omega_1^+ -\omega_2^- - \omega_3^- $. When considering long-term dynamics, the oscillatory term in~\eqref{eq_corr3_solved} can be neglected, hence~\eqref{eq_corr3_solved} becomes
\begin{equation}
\label{J_123}
J_{1,2,3}=\frac{2V_{1,2,3}^{(2)*}(n_1^+ n_2^- + n_1^+n_3^- -n_2^-n_3^-)}{\omega_1^+ -\omega_2^- - \omega_3^- + i\delta(\Delta\omega)},
\end{equation}
where we add $i\delta(\Delta\omega)$ in the denominator to avoid a divergent quantity in case the resonance conditions apply.
Note that
\begin{equation}
\label{delta}
\Im\{[\Delta\omega + i\delta(\Delta\omega)]^{-1}\} =-\delta(\Delta\omega).
\end{equation}
Combining~\eqref{eq_2nd_order_corr},~\eqref{corrIII} and~\eqref{J_123}, we obtain a time evolution equation of $n_1^+$.
Proceeding from the second part of (\ref{longterm}), we get a time evolution equation of $n_1^-$ in a similar way. The time evolution equations for $n_1^{\pm}$ can be written as
\begin{subequations}
	\begin{eqnarray}
	\frac{dn_1^+}{dt}&=&4\int_0^{\pi/a} \vert V_{1,2,3}^{(2)}  \vert^2
	n_1^+ n_2^- n_3^- \Bigl( \frac{1}{n_1^+}-\frac{1}{n_2^-} -\frac{1}{n_3^-}\Bigr)
	\delta_{1,2+3}\, \delta_{\omega_1^+,\, \omega_2^- + \omega_3^-}dk_{23} \label{27a}\\
	\frac{dn_1^-}{dt}&=&8\int_0^{\pi/a}  \vert V_{2,1,3}^{(2)}  \vert^2
	n_1^- n_2^+ n_3^- \Bigl( \frac{1}{n_1^-}-\frac{1}{n_2^+} +\frac{1}{n_3^-}\Bigr)
	\delta_{1,2-3}\, \delta_{\omega_1^-,\, \omega_2^+ - \omega_3^-}dk_{23}\label{27b}
	\end{eqnarray}
\end{subequations}

\onecolumngrid

\end{document}